\documentclass[useAMS,usenatbib,usegraphicx]{mnras}
\usepackage{graphicx,amssymb,amsmath,times,verbatim}
\usepackage[normalem]{ulem}
\def\lta{\mathrel{\spose{\lower 3pt\hbox{$\mathchar"218$}}
     \raise 2.0pt\hbox{$\mathchar"13C$}}}
\def\gta{\mathrel{\spose{\lower 3pt\hbox{$\mathchar"218$}}
     \raise 2.0pt\hbox{$\mathchar"13E$}}}

\def\mathnew{\mathsurround=0pt}

\def\simov#1#2{\lower .5pt\vbox{\baselineskip0pt \lineskip-.5pt
\ialign{$\mathnew#1\hfil##\hfil$\crcr#2\crcr\sim\crcr}}}

\def\simless{\mathrel{\mathpalette\simov <}}

\title[OJ 287 $\gamma$-ray QPO]{A Possible $\gamma$-ray Quasi-periodic Oscillation
of $\sim$ 314 days in the Blazar OJ 287}

\author[Kushwaha et al.]{Pankaj Kushwaha$^{1,2}$\thanks{E-mail: pankaj.tifr@gmail.com}
\thanks{Aryabhatta Postdoctoral Fellow},
Arkadipta Sarkar$^3$\thanks{E-mail: sarkadipta@gmail.com}, Alok C. Gupta$^{1}$\thanks{E-mail: acgupta30@gmail.com}, Ashutosh Tripathi$^{4}$,
Paul J. Wiita$^{5}$\\
\\
$^{1}$Aryabhatta Research Institute of Observational Sciences (ARIES), Manora Peak, Nainital 263002, India \\
$^{2}$Department of Astronomy (IAG-USP), University of Sao Paulo, Sao Paulo 05508-090, Brazil \\
$^{3}$Department of High Energy Physics, Tata Institute of Fundamental Research,
Mumbai 400005, India\\
$^{4}$Center for Field Theory and Particle Physics and Department of Physics, Fudan University, 220 Handan Rd., Shanghai 200433, China\\
$^{5}$Department of Physics, The College of New Jersey, P.O.\ Box 7718, Ewing, NJ 08628-0718, USA   \\
}

\begin{document}
%   \linenumbers

\maketitle

\begin{abstract}
We report the detection of a probable $\gamma$-ray quasi-periodic oscillation
(QPO) of around 314 days in the monthly binned 0.1 -- 300 GeV $\gamma$-ray {\it
Fermi}-LAT light curve of the well known BL Lac blazar OJ 287. To identify and quantify
the QPO nature of the $\gamma$-ray light curve of OJ 287, we used the Lomb-Scargle
periodogram (LSP), REDFIT, and weighted wavelet z-transform (WWZ) analyses. We
briefly discuss possible emission models for radio-loud active galactic nuclei (AGN)
that can explain a $\gamma$-ray QPO of such a period in a blazar. Reports of changes
in the position of quasi-stationary radio knots over a yearly timescale as well as
a strong correlation between gamma-ray and mm-radio emission in previous studies
indicate that the signal is probably associated with these knots.
\end{abstract}


\section{Introduction} \label{sec:intro}
Blazars are extremely variable active galactic nuclei (AGNs) with overall emission
dominated by a relativistic jet that is aligned at close angles to our line of sight.
They have been found to exhibit large amplitude flux variability spanning the entire
accessible electromagnetic spectrum on all time scales feasible by the current observing
facilities and available data \citep[e.g.][]{2018ApJ...863..175G,2013A&A...559A..20H,
1988ApJ...325..628S,1973ApJ...179..721V}. The observed flux variability is generally
stochastic and broadly shares the scale-free statistical properties exhibited by
accretion-powered sources \citep{2016ApJ...822L..13K,2017ApJ...849..138K}.
So, regardless of the underlying emission mechanisms and detailed physical attributes
this implies that accretion physics is the main driver.

Flux variability on diverse timescales is one of the hallmark properties of
accretion-powered sources. In fact, based on statistical similarities of flux variations,
AGNs have been claimed to be a scaled-up version of Galactic black hole X-ray binaries
(BHXBs) \citep[e.g.][and references therein]{2015SciA....1E0686S}. Yet detection
of quasi-periodic oscillations (QPOs) is rare in blazars, while it is a relatively
common phenomena in BHXBs. Further, almost all the detected QPOs in BHXBs are
transient, and widely believed to be related to their accretion (disc) 
\citep{1997MNRAS.292..679L,2006MNRAS.367..801A,2015SciA....1E0686S}.
On the contrary, in AGNs, QPOs have been observed to be both transient \citep[e.g.]
[]{1985Natur.314..146C,1973ApJ...179..721V,1985Natur.314..148V,2013MNRAS.434.3122P,
2009ApJ...690..216G,2009A&A...506L..17L}, which constitute
the majority of the reported claims, as well as persistent \citep[e.g. ][]{1988ApJ...325..628S,2013A&A...559A..20H,2018ApJ...854...11T}. AGN, being accretion powered,
transient putative QPOs are expected
naturally based on the similarity with BHXBs, while the apparently permanent
ones can be due to jet precession and/or be a prime signal of the host being a supermassive
BH (SMBH) binary \citep[e.g. ][]{1988ApJ...325..628S,2018ApJ...854...11T}, as expected
in the hierarchical model of cosmological evolution. Thus, exploration of such signals
in blazar time series is a key approach to investigate the still not fully understood 
accretion physics and the connection between central engine, accretion, and jet,
in addition to searching for SMBH binaries.

Searches for periodicity in blazars and AGNs generally have been very challenging,
mainly due to limitations associated with operation of observing facilities leading
to gaps in the data and the huge range of energy dependent temporal variability
shown by these sources \citep[e.g. ][]{2013MNRAS.434.3122P}. To date, QPOs have
been reported in only a handful of blazars and these span a huge range of timescales:
from a few tens of minutes \citep[e.g.][]{1973ApJ...179..721V,1985Natur.314..148V},
through a few hours \citep[e.g.][]{2009ApJ...690..216G}, and months \citep[e.g.][]{2013MNRAS.434.3122P,
2016AJ....151...54S,2019MNRAS.484.5785G} to decades \citep{1988ApJ...325..628S} and
in different energy
bands. The majority of these detection claims are marginal as of now \citep[e.g]
[and references therein]{2019MNRAS.484.5785G}. With AGN variability being
stochastic and dominated by red/flicker-noise, searches for QPO signals are fraught
with risks and demand high quality data and  thus is still an evolving research field.

OJ 287 \citep[$z=0.306$;][]{1985PASP...97.1158S} is a BL Lacartae (BL Lac) object --
a subclass of blazar characterized by very weak or complete absence of emission
lines in their optical spectra. The source is famously known for its double-peaked 
outbursts in optical bands which repeat every $\sim$12-yr \citep{1988ApJ...325..628S}.
Ever since the discovery of OJ 287 \citep[VRO 20.08.01;][]{1967AJ.....72..757D}
and the identification of an optical counterpart in 1970 \citep{1970ApL.....6..201B},
 it has been
a prime target of observations, owing to a variety of observationally favorable
properties, including high and correlated multi-wavelength activities \citep[and
references therein]{1970ApL.....6..201B,1985PASP...97.1158S}. In fact, a 39.2 minute QPO signal was reported by \citet{1973ApJ...179..721V} in one of the
earliest targeted optical observations of the source \citep[see also][]{1974MNRAS.168..417F}.
They also successfully extracted its historic optical light curve from records
dating back to 1894 and found four high activity periods, each lasting over
several months.

A wide range of observational features in different energy bands have made OJ 287
an ideal astrophysical laboratory to probe a complete spectrum of problems pertaining
to blazars, ranging from jet dynamics \citep[and references therein]{2018MNRAS.478.3199B, 2019arXiv190403357Q}, high energy emission mechanisms \citep[and references therein]{2020Galax...8...15K,2018MNRAS.479.1672K,
2018MNRAS.473.1145K,2013MNRAS.433.2380K,2018bhcb.confE..22K} to the general theory
of relativity \citep[and references therein]{2018ApJ...866...11D}. Extensive and
intensive temporal investigations on a wide range of timescales has led to claims
of a diverse range of QPOs in OJ 287  ranging from a few tens of minutes \citep[e.g.]
[]{1973ApJ...179..721V,1985Natur.314..148V}, to a few months \citep[e.g.][]
{2013MNRAS.434.3122P}, to years \citep[e.g.][]{2016BaltA..25..237D,2016AJ....151...54S,
2018MNRAS.478.3199B} to decades \citep[e.g.][]{1988ApJ...325..628S} in various
bands of the EM spectrum. In many cases, studies have reported  conflicting/different
results for observations during similar periods. For example, \citet{1985Natur.314..148V} 
reported of a \(15\)-min QPO in radio and optical for 1981 -- 1983 while
\citet{1985Natur.314..146C} reported a roughly \(23\)-min QPO in the optical in 1983
-- 1984. The same was the case with the \(40\)-min QPO in the optical
reported by \citet{1973ApJ...179..721V} during 1972 which was also found by 
\citet{1974MNRAS.168..417F} but not by \citet{1974ApJ...191L.109K}. To the best
of our knowledge, OJ 287 is the blazar (and AGN) for which the highest number of QPO
claims have been made.

Here, we present our work on a search for possible QPOs in the 9.5-yr (August 5,
2008 -- February 5, 2018) long \emph{Fermi}-LAT $\gamma$-ray light curve of OJ 287
derived by binning it into  30-day intervals. One of our motivations has been the
finding of a probable yearly timescale periodicity of the location of quasi-stationary
radio knots \citep{2018MNRAS.478.3199B} and the correlation between $\gamma$-ray emission
and these knots \citep{2011ApJ...726L..13A}. In \S2, we summarize the data
and the reduction procedure used to extract the light curve.
\S\ref{sec:analysis} presents the results from different methods of QPO analyses,
followed by our discussion, inferences, and conclusions in \S\ref{sec:discussion}.

\section{Fermi $\gamma$-ray Data} \label{sec:data}
The Large Area Telescope (LAT) onboard  the space based \emph{Fermi} observatory is
currently the best sensitive facility in the MeV-GeV ($>$20 MeV) energy range, continuously
surveying the sky every 90 minutes since August 2008 \citep{2009ApJ...697.1071A}.
Though it is capable of detecting photon events between 20 MeV to $>$ TeV energies,
a large point spread function (PSF) at low energies ($<$100 MeV) and uncertainty
in photon-particle events separations make those low energy results unreliable,\footnote{There is confusion
as to whether those detected photons are from the source of interest or due to many sources
within the PSF or a particle event tagged as a photon event.} while the low event statistics at
$>300$ GeV make short term studies infeasible in that range.

For extracting the light curve, we used the \textsc{Fermitool} (version 1.0.1) and
followed the standard data reduction procedure. We used the LAT data processed
with the PASS8 ($P8R2$) instrument
response function and selected the ``SOURCE'' class events from a 
circular region of 15$^\circ$ centered on the source with energies between
100 MeV to 300 GeV. At the same time, a zenith angle cut of 90$^\circ$ was employed
to avoid $\gamma$-ray confusion from the Earth's limb. We used the time filter
expression ``$(DATA\_QUAL>0)\&\&(LAT\_CONFIG==1)$'', the standard prescription for
generating the good time intervals. Following this, we calculated the
exposure for the events on the adopted circular source region plus an additional
10$^\circ$ annulus around it. 

\begin{figure}%[h]
\begin{center}
\includegraphics[scale=0.65]{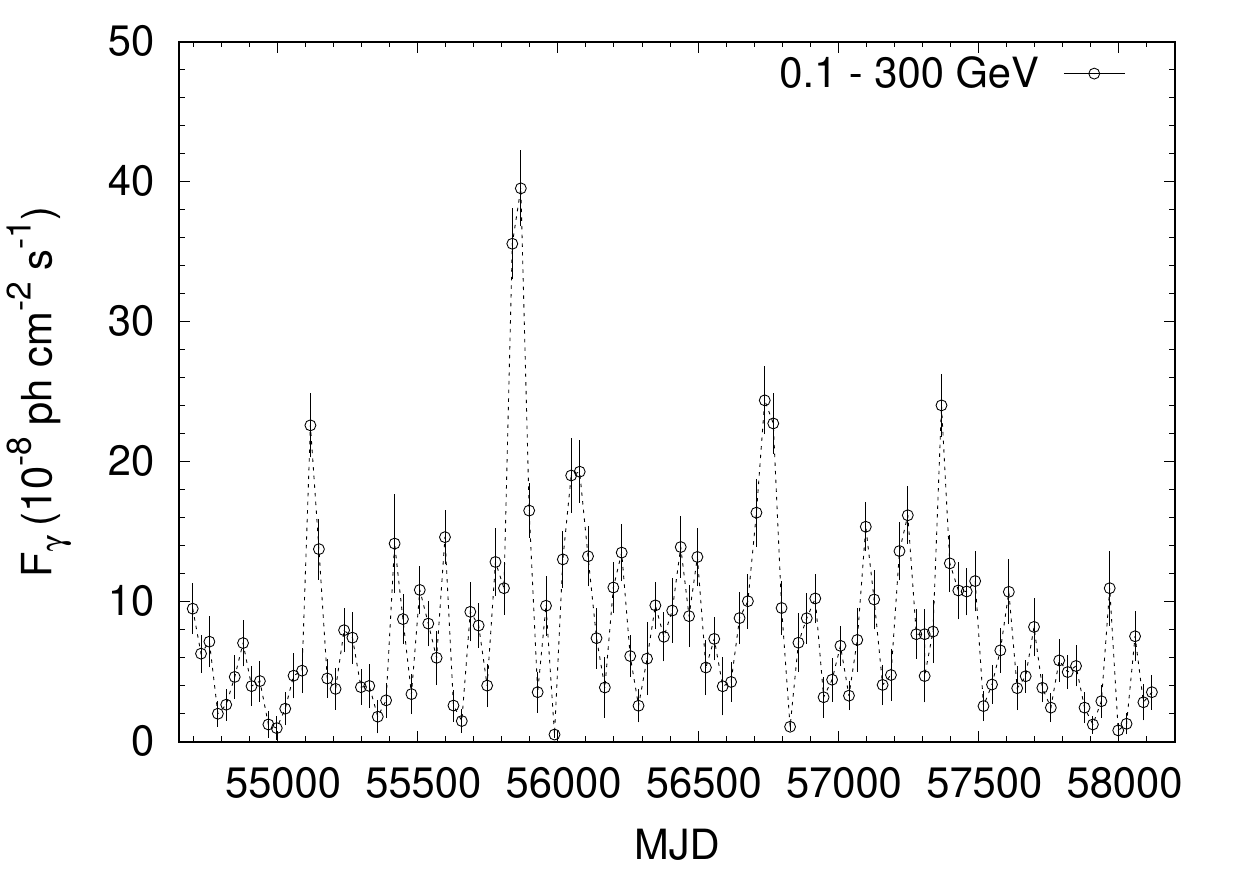}
\end{center}
\caption{OJ 287 $\gamma$-ray light curve extracted from a 30-day binning of \emph{Fermi}-LAT
data between August 2008 and February 2018 (MJD: 57683 -- 58153).}
\label{fig:lc}
\end{figure}

The photon flux was extracted through unbinned likelihood analysis provided within
the software using an input spectral model XML file. The input XML model file  is comprised
of the Galactic (gll\_iem\_v06) and isotropic extra-galactic (iso\_P8R2\_SOURCE\_V6\_v06) contributions though their respective templates and
all the point sources within the angular extension covered by the exposure map. It was
generated using the 3rd \emph{Fermi}-LAT source catalog \citep[3FGL; ][]{2015ApJS..218...23A}
in which the source spectrum has been described as a log-parabola. The fit was performed
iteratively by removing sources contributing insignificantly, measured by a Test
Statistic (TS) value of $\simless$ 0 until the optimization converged \citep[e.g.][]{2014ApJ...796...61K}. The 0.1-300 GeV photon flux
light curve of the source, after applying a TS cut of 9.0 --- equivalent to a $\sim$
3$\sigma$ detection --- is shown in Figure \ref{fig:lc}.

\section{QPO Analysis and Estimation}\label{sec:analysis}

We employed three different methods to search for periodicities: Lomb-Scargle periodogram (LSP); REDFIT analysis;
and wavelet z-transform (WWZ),  We used their basic
underlying statistics to access the significance of features. The LSP and REDFIT
are both global, in the sense that they treat the whole time series in a go, while
the wavelet z-transform is local and looks for periodicities around each of the
data points.

\subsection{Lomb-Scargle Periodogram}\label{subsec:LS}

The Lomb-Scargle (L-S) periodogram is one of the basic and most widely used methodologies
in  time series analyses to look for periodicities \citep{1976Ap&SS..39..447L,
1982ApJ...263..835S}, primarily due to its ability to handle non-uniformly sampled
data, which are the norm in astronomical observations \citep{1981AJ.....86..619F}. The
method is basically a projection of the time series on sinusoidal functions and
is equivalent to a $\chi^2$ fit statistic for uniformly sampled data and a weighted
$\chi^2$ fit statistic for non-uniformly sampled data \citep{1981AJ.....86..619F}. 
The statistics then in turn allow assessment of the significance of features as signals
of interest if any are present. For more details, see \citet{2019MNRAS.484.5785G} and
references therein.

The right panel of Fig. \ref{fig:fig2} shows the normalized power of the Lomb-Scargle
periodogram for the extracted {\it Fermi}-LAT data. The dotted line indicates
the 99.99\% significance level which corresponds to the false alarm probability (FAP) of {0.0001}.
However, this FAP significance estimate strictly holds only for Gaussian type random noise,
while AGNs show red-noise variability. Thus, we estimated the significance against the
red-noise spectrum of the observed light curve by simulating 1000 light curves (see \S
\ref{subsec:sig}), shown by the blue dashed curve in the right panel of Figure \ref{fig:fig2}.
The analysis revealed a signal centred at \(314\) days with a significance of \(3\sigma\).
This signal could be a QPO, but it requires support from other methods.

\subsection{Wavelet Analysis} \label{subsec:wzt}
Wavelet analysis is a powerful technique which is used for searching the periodicities
in the data by decomposing the signal into frequency-time space simultaneously. In
other words, this technique measures the non-stationarity of the dataset and indicates the temporal
extent of any possible periodic feature. For more
details on our approach see \citet[and references therein]{2018A&A...616L...6G}.
 Here, we used the
WWZ\footnote{https: //www.aavso.org/software-directory} software which uses the
wavelet z-transform method to search and assess the significance  of any periodicities
present in the data  \citep[e.g.][and references therein]{2013MNRAS.436L.114K,
2016ApJ...832...47B,2019MNRAS.487.3990B,2017ApJ...835..260Z,2018ApJ...853..193Z}. 

\begin{figure}%[ht]
  \centering
 \includegraphics[scale=0.30]{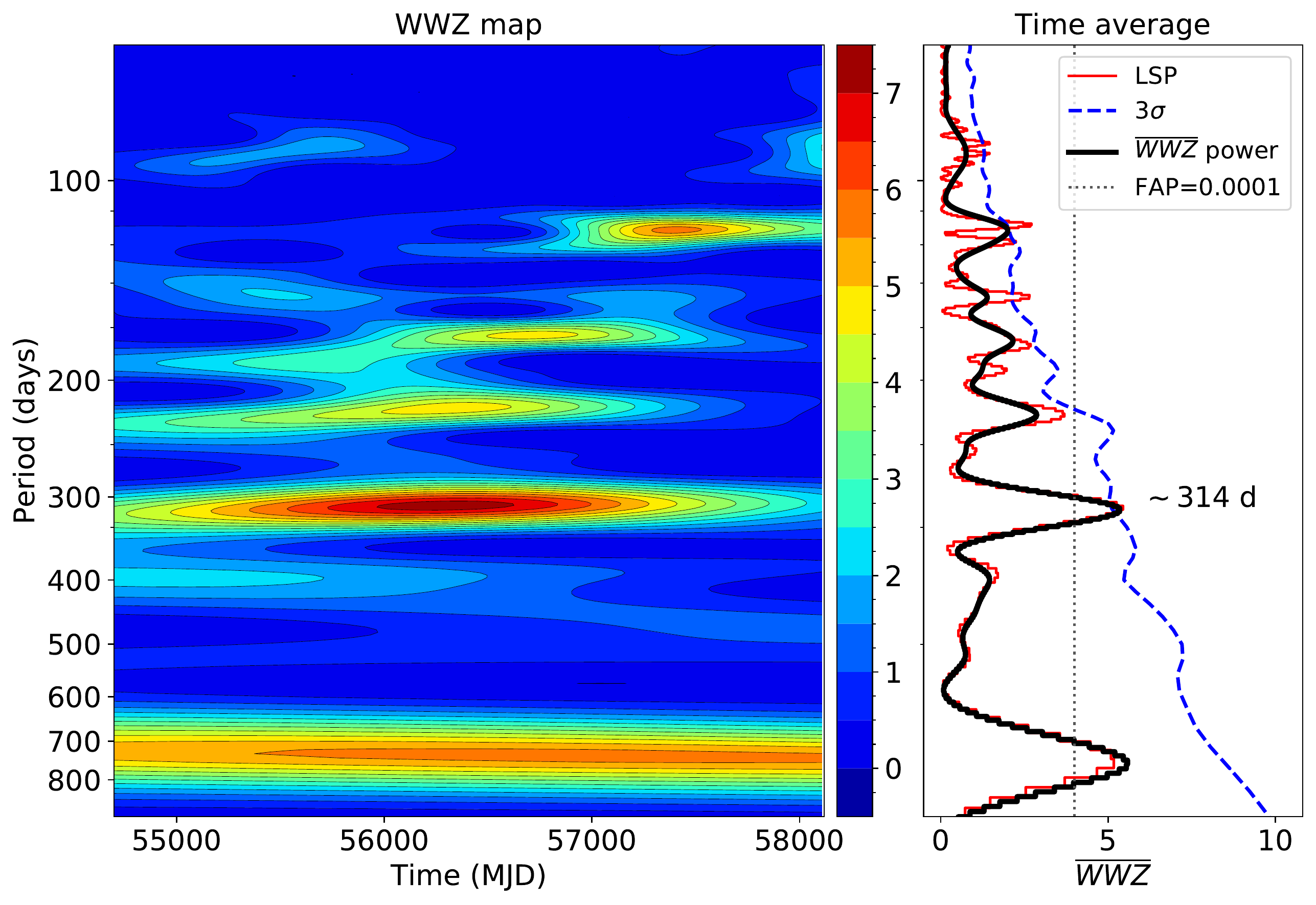}
\caption{LSP and WWZ results of the OJ 287 $\gamma$-ray time series data (see Sections
\ref{subsec:LS} and \ref{subsec:wzt}). The left panel is the full wavelet result and the
right panel shows the averaged WWZ power (black) and the LSP results (red). The dotted vertical line in
the right panel corresponds to FAP=0.0001 (99.99\%) while the blue dashed curve shows the 
\(3\sigma\) significance against the red-noise spectrum of the observed light curve
(see \S\ref{subsec:sig}).}
\label{fig:fig2}
\end{figure}

The left panel of Fig. \ref{fig:fig2} shows a color density plot of the  WWZ power
for the OJ~287 $\gamma$-ray data as a function of both time and frequency. The right
panel shows the time averaged WWZ power as a function of frequency. The WWZ
method also shows a highly significant \(3\sigma\) signal at \( 314\) days,
corroborating the LS result. Further, the signal is persistent for most of the time
duration.

\subsection{REDFIT Analysis} \label{subsec:redFit}

As already mentioned, the light curves of AGNs are generally dominated by red-noise
which originates due to some stochastic processes in the accretion disc, or jet,
in the case of blazars \citep[e.g.,][and references therein]{2014ApJS..213...26F,
2017ApJS..229...21X,2018AJ....155...31H}. 
Red-noise spectra are characteristic of auto-regressive (AR) processes
where activity at an instant is related to past activities \citep[e.g.][]{
2002CG.....28..421S, 2014ApJS..213...26F,2018ApJ...853..193Z,2017ApJS..229...21X, 2018AJ....155...31H,
2018A&A...616L...6G}. AGN emissions are generally autoregressive and usually can be convincingly
modeled using an AR1 process where the present emission depends on the emission that
immediately precedes it, i.e.,
  \begin{equation}
    \mathcal{F}\left(t_{i}\right)=A_{i} \mathcal{F}\left(t_{i-1}\right)+\sqrt{1-A^{2}} \epsilon\left(t_{i}\right) ,
  \end{equation}
  where $\mathcal{F}\left(t_{i}\right)$ is the flux at time $t_i$ and $A_{i}=
  \exp ([t_{i-1}-t_{i}] / \tau) \in[0,1]$, $A$ is the average autocorrelation coefficient,
  $\tau$ is the timescale of the autoregressive 
  process, and $\epsilon$ is a Gaussian distributed random variable with zero mean and
  unit variance. The theoretical power spectrum of an AR1 process is given by \citep{2002CG.....28..421S}
  \begin{equation}
    \label{eq:ar1_spec}
      G_{r r}\left(f_{j}\right)=G_{0} \frac{1-A^{2}}{1-2 A \cos \left(\pi f_{j} / f_{N y q}\right)+A^{2}} ,
    \end{equation}
where $G_0$ is the average spectral amplitude, $f_i$ are the frequency points
and $f_{Nyq}$ is the Nyquist frequency. The code REDFIT \citep{2002CG.....28..421S} estimates the spectrum
using LSP in combination with Welch-Overlapped-Segment-Averaging. The resulting
spectrum is then modeled using Equation \ref{eq:ar1_spec} and calculates the significance
of the powers in the periodogram.

In the REDFIT method, \(f_{Nyq} = H_{fac}/(2\Delta t)\) where the additional factor
$H_{fac}$ is to avoid the noisy high frequency end of the spectrum affecting the
fit given by equation \ref{eq:ar1_spec}. This factor can be found through the non-parametric test
employed in REDFIT to check the consistency of the theoretical spectrum with that of
the data spectrum. The value of \(G_0\), on the other hand, is determined
by demanding that area under the theoretical spectrum is same as the variance of the
data.

\begin{figure}%[ht]
\centering
 \includegraphics[scale=0.35]{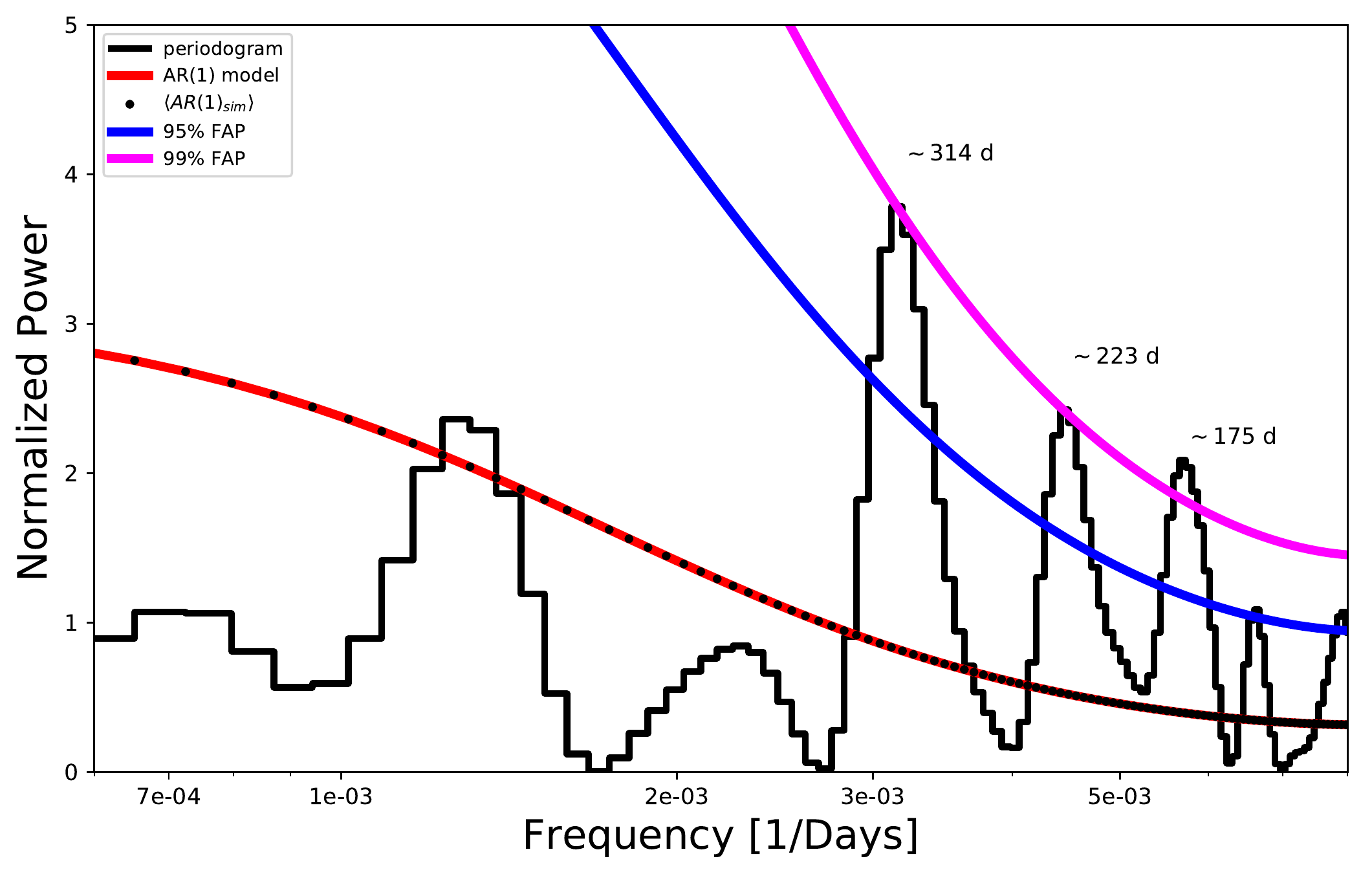}
\caption{OJ 287 power spectrum and its significance using the REDFIT method 
(\ref{subsec:redFit}). The red, blue, and magenta curves respectively represent
the AR1 curve and the 95\% and 99\% significance levels. The black
dotted points (seen within the red curve) are the mean simulated AR1 spectrum.}
\label{fig:fig3}
\end{figure}

Both the LS and WWZ methods show a signal of \(\sim314\) days, with the WWZ result
revealing that most of the power of this signal is seen during the central portion of
the time series.  Hence, we used a triangular window in the REDFIT method. Running 
REDFIT resulted in \(G_0 = 4.97\times10^{-13}, A = 0.52\), while the statistical
consistency of theoretical AR1 spectrum and data spectrum resulted in \(H_{fac} = 0.5~ 
(\Delta t = 30)\). The resulting bias-corrected power spectrum (black) is shown
along with the AR1 spectrum (red) in Fig. \ref{fig:fig3}. The blue and magenta
curves correspond to significance levels
of 95\% and 99\%, respectively, derived from the simulation of 2000
instances of the best-fit spectrum given by Equation \ref{eq:ar1_spec}. The bias
corrected mean of these simulated spectra is also over-plotted above the AR1 spectrum
(red) for comparison. 

The method suggests three time scales with significance \(\gtrsim\)99\%. The lowest
frequency peak corresponds to a period of about \(314\) d, coinciding
with the QPO revealed by LS and WWZ. The subsequent higher frequencies correspond to
periods of \(223\) d and \(175\) d, respectively, but these are without any counterparts
in the LSP and WWZ analyses. Given that \citet{2016AJ....151...54S} have reported a
\(\sim438\) d period in near-infrared (NIR) K-band data of OJ 287, the 
\(223\) d feature could be a harmonic of that component while the \(175\) d signal
seems to be an alias of the \(314\) d QPO identified by all the methods employed
here. It should, however, be noted that a \(438\) d period claimed by
\citet{2016AJ....151...54S} is not significant in the REDFIT output -- neither in
the {\it Fermi}-LAT $\gamma$-ray light curve nor in the optical R-band light curve.
To further understand this discrepancy, we checked the coverage of the SMARTS K band
data on OJ 287 and found that it has a very poor coverage, at $<20\%$ of the
considered duration. 
As both LAT and R band are sampled better than the NIR K-band, we believe
the claimed signal to be an outcome related to data sampling. Further, 
\citet{2016AJ....151...54S} have used a 2\(\sigma\)
threshold for considering LAT data points for time seeries analysis compared to the more commonly accepted 3\(\sigma\) criterion.

\subsection{Significance estimation} \label{subsec:sig}
Two independent methods were used for the estimation of the significance of the
dominant period. The first method, which is employed by REDFIT, is similar to that
of \cite{Vaughan2005} where the significance is estimated from the $\chi^2$ distribution
of periodogram points about the model. However, contrary to \cite{Vaughan2005},
where the assumed model is a power-law, REDFIT uses Equation \ref{eq:ar1_spec}, thereby
preventing underestimation of PSD peak significances, especially at low frequencies.

Along with the above analytical method for significance estimation, a Monte-Carlo 
method was also used to get the significance of the LSP as well as WWZ peaks. One
thousand light curves, having the same PSD and flux distribution, were simulated
using the approach of \cite{Emmanoupoulos2013}. From the distribution of PSD power
at each frequency, the $3\sigma$ significance level was estimated.

\begin{figure}%[ht]
\centering
 \includegraphics[scale=0.35]{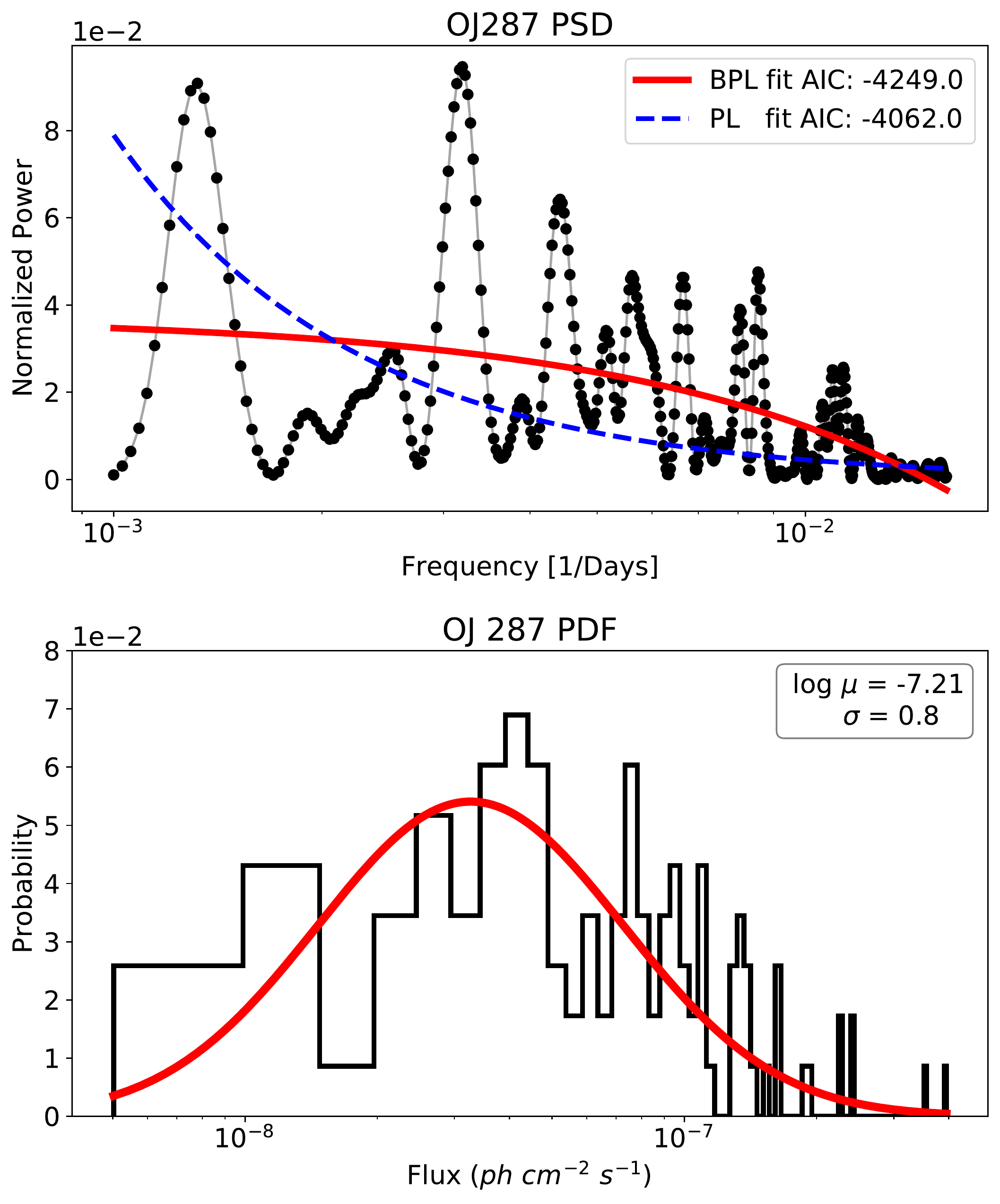}
\caption{{\it Top}: Power spectrum of the observed light curve. The blue and the red
curve represent a bending power-law and power-law fit to the power spectrum with AIC
value also given. {\it Bottom}: Observed flux distribution of the source light curve.
The red curve is the best-fit log-normal used for simulating artificial time series
to estimate the significance of the signal (see \ref{subsec:sig}).}
\label{fig:fig4}
\end{figure}

To determine the PSD model for generating artificial time series used for significance
estimation as mentioned above, we performed maximum-likelihood estimates of the data PSD.
We modeled the data PSD both with a power-law and a bending power-law model given by
\begin{equation}\label{eq:bPL}
P(\nu) = \frac{\nu^{-\alpha_{low}}}{1 + (\nu/\nu_{b})^{\alpha_{high} - \alpha_{low}}}, 
\end{equation}
where $\alpha_{low}$ and $\alpha_{high}$ are the two spectral indices and $\nu_b$
is the bending frequency. The Akaike information criterion (AIC) was used to determine
the best-fit model. The fitted models and the corresponding AIC values are shown in the
top panel of Figure \ref{fig:fig4}. As per the AIC value, a bending power-law,
with {\it parameter values of} $\alpha_{low} = 0.52$, $\alpha_{high} =1.39$ and $\nu_b = 0.015\ d^{-1}$
is favored over the power-law. For the flux distribution, a log-normal distribution,
as shown in the bottom panel of Figure \ref{fig:fig4}, was found to describe it well.

We further verified the detected QPO duration by extracting {\it Fermi}-LAT light curves with
binnings of 21 d and 15 d. Analyses of both of these light curves using the above mentioned
methods and procedures also show the \(314\) d signal, though the significance
is slightly lower in the 15 d bin light curve ($> 0.95$).

\section{Discussion and Conclusions}\label{sec:discussion}

We found a highly probable QPO of about 314 days in the 9.5 yr long $\gamma$-ray
light curve of OJ 287, extracted from 30-day binning of the \emph{Fermi}-LAT survey
mode data. The signal is consistently identified by all the three methods commonly
used to explore time series of blazars/AGNs and in astronomy in general. Further,
examination of the wavelet z-transform (Figure \ref{fig:fig2}) shows that the signal
is clearly present at least for about 8 cycles and seems to weakened/disappeared after MJD 57400, strongly indicating that it is
not a manifestation of a stochastic feature associated with a flicker/red-noise time
series. 
%{\bf Our data also show that this weakening/disappearance is coincident with
%the spectral hardening of MeV-GeV emission, indicating it to be related to spectral
%changes \citep[e.g. see also][]{2018MNRAS.479.1672K}}. 
It should be noted that for broadband,
photon statistics limited observing
facilities/instruments such as \emph{Fermi}-LAT here, the observed photon flux is
highly dependent on the spectral state of the source and a change in spectral state
can mask a real change in photon flux in case the increment is mainly at the high
end of the instrument-energy-band \citep[e.g. ][]{2018MNRAS.479.1672K}. In such
a case, even if the signal is present, it will not be reflected in the observed photon
flux \citep[see][]{2018bhcb.confE..22K} and thus will be missed by the analysis method.
Further, even though \emph{Fermi}-LAT survey data in principle
allow for temporal variation studies on time-scales as short as 1.5 hours
and even less, none of the currently known \emph{Fermi}-LAT AGN sources are bright
enough to allow a high quality, homogeneously sampled continuous light curve on those
short timescales over the \emph{Fermi}-LAT operation period. Given that blazar
variability is essentially stochastic, a high quality, continuous, and homogeneously
sampled data train is required to negate the artifacts associated with gaps in the
data. So the challenges associated with finding QPO signals in these time series
are major, despite substantial methodological progress in the field.

Since OJ 287 is a BL Lac type of  blazar, where accretion provides the ultimate
source of power but boosted jet radiation dominates the entire emission, such a
QPO signal can be a manifestation of many aspects of interactions between the entities
that constitute an AGN and affect jet emission. These include frequently suggested
accretion disc scenarios \citep[e.g.][and references therein]{2013MNRAS.434.3122P,
2016ApJ...832...47B}, a precessing jet due to gravitational interaction with another
massive object \citep{2018MNRAS.478.3199B,2019arXiv190403357Q}, binary SMBH dynamical
interaction scenarios \citep[e.g.][and references therein]{1988ApJ...325..628S,
2018ApJ...866...11D}, a jet scenario driven by plasma instabilities or turbulence
 \citep{2014ApJ...780...87M,2016ApJ...820...12P}, or a combination of these.
Though BL Lacs in general are believed to have usually undetectable accretion disc
contributions to their observed emission within our current level of observational
capabilities, some of these have been inferred to have an
active, dynamic accretion disc \citep[e.g. OJ 287;][]{1988ApJ...325..628S,
2018MNRAS.473.1145K}. For OJ 287, it has been argued to be a dynamic system
comprised of a pair of SMBHs \citep[and references therein]{2018ApJ...866...11D}.

QPOs of diverse timescales have been reported in OJ 287 in different energy bands.
In optical bands, a 39.2-minute periodicity was suggested in 1972 observations
\citep{1973ApJ...179..721V} and a  23-minute periodicity was claimed in 1983
observations \citep{1985Natur.314..146C}. Some evidence of possible optical quasi-periods 
of a few hours and $\sim$ 50 days \citep{2013MNRAS.434.3122P}, 400 days
\citep{2016ApJ...832...47B}, and a persistent $\sim$ 12 yrs QPO
\citep{1996A&A...305L..17S} have also been reported. The 12-yr QPO is 
most frequently attributed to the source being a binary SMBH system with the temporal
signal resulting from the impact of the secondary SMBH on the accretion disc of the 
primary \citep[but see \citet{2018MNRAS.478.3199B,2019arXiv190403357Q}]
{1988ApJ...325..628S,2008Natur.452..851V}. This model has been reasonably successful
in predicting the timing of subsequent flares \citep{2018ApJ...866...11D}.
On the other hand, almost all optical QPOs on timescales of year and shorter have
been argued to be associated, one way or another, with accretion disc fluctuations,
either directly \citep[e.g.][]{2016ApJ...832...47B,2013MNRAS.434.3122P} and/or through
a disc modulation imprinted onto the jet with relativistic effects shortening the
apparent period \citep[e.g.][]{2012NewA...17....8G}. The accretion disc origin for
optical QPOs is primarily motivated from observation of QPOs in accretion-powered
sources and observational (big blue bump) support for SMBH/AGN disc emission being
in optical-UV bands.

As is the case in optical bands, several possible QPOs have been reported in radio
bands as well, e.g. a 15.7-minute periodicity by \citet{1985Natur.314..148V} in the
observations taken 1981 at 37 GHz frequency, while indications of QPOs of periods
from about a year up to 7.5 years have been reported by \citet{1998ApJ...503..662H}
and \citet{2016BaltA..25..237D} at 4.8 GHz, 8.0 GHz, and 14.5 GHz. Since emission
at radio frequencies arises from the jet, the radio QPOs have been argued to be
either due to dynamics in the jet such as
shocks \citep[e.g.][]{1998ApJ...503..662H} or precession \citep[e.g][]
{2018MNRAS.478.3199B,2019arXiv190403357Q}. In fact, nearly periodic variability is not
limited to flux but also seems to be present in the positional location of quasi-stationary
VLBA features (called knots) seen in radio images at 15 GHz \citep{2018MNRAS.478.3199B}.
The quasi-stationary features seem to vary over roughly 1 yr periods while the moving
VLBA features suggest 22 yr periodic variations. Based on these 
observations, \citet{2018MNRAS.478.3199B} have argued OJ 287 to be a
system with a precessing and rotating jet  \citep[see also][]{2019arXiv190403357Q}.

For the QPO signal of $\sim$ 314 days reported here in the $\gamma$-ray band of
OJ 287, both accretion disc based variations, through their imprint on the jet, as
well as a purely jet based origin seem plausible. In the disc-imprint-on-the-jet
based interpretation, a tempting possibility is that the roughly 12-yr optical QPO
signal, attributed to the impact of the secondary SMBH on the accretion disc of the
primary, could manifest on timescales of less than a year by  temporal compression
by the Doppler factor ($\sim$14) inferred for the jet \citep{2011ApJ...726L..13A}.
However, jet emission being broadband, this also implies expected QPO signals across
many electromagnetic bands with similar temporal profiles unless there is strong
spectral change, such as the observed MeV-GeV spectral changes during its first
detected VHE activity \citep{2018MNRAS.479.1672K}. Unfortunately, temporal coverage 
in other bands is rather too sparse to either pin-point or rule out this possibility.
Given a totally jet-dominated emission in OJ 287, a sub-yearly modulation of the
quasi-stationary jet features/knots \citep{2018MNRAS.478.3199B} and the association
of $\gamma$-rays with these knots \citep{2011ApJ...726L..13A}, a precessing jet
seems to provide a probable mechanism for the observed $\gamma$-ray QPO.

\section*{Acknowledgements}
 The authors thank the referee for constructive comments and suggestions. 
PK acknowledges support from A-PDF grant (AO/A-PDF/770) and partial support from
FAPESP grant no.\ 2015/13933-0. AS acknowledges support of the Department of Atomic Energy, Government of India, under project no. 12-R\&D-TFR-5.02-0200.

The \emph{Fermi}-LAT  Collaboration  acknowledges support  for LAT development,
operation and data analysis from NASAand   DOE   (United   States),   CEA/Irfu
and IN2P3/CNRS(France),  ASI  and  INFN  (Italy),  MEXT,  KEK,  and  JAXA(Japan),
and the KA  Wallenberg  Foundation,  the  SwedishResearch Council and the National
Space Board (Sweden).Science analysis support in the operations phase from INAF 
(Italy) and CNES (France) is also gratefully acknowledged. 

\section*{Data Availability}
The datasets were derived from sources in the public domain: [Fermi-LAT,
https://fermi.gsfc.nasa.gov/cgi-bin/ssc/LAT/LATDataQuery.cgi]
The derived data generated in this research will be shared on reasonable request
to the corresponding author.

\end{document}